\documentclass[a4paper, 11pt]{article}
\usepackage{amsmath}
\usepackage{graphicx}
\usepackage{latexsym}

\def\b{\begin{eqnarray}}
\def\e{\end{eqnarray}}
\def\n{\noindent}

\begin{document}

\begin{center}

{\Large\textbf{Equations of the Camassa-Holm Hierarchy
\\}} \vspace {5mm}  \noindent

{\large \bf Rossen I. Ivanov} \footnote{School of Mathematical
Sciences, Dublin Institute of Technology, Kevin Street, Dublin 8,
Ireland, Tel:  + 353 - 1 - 402 4845,   Fax:  + 353 - 1- 402 4994,
e-mail: rivanov@dit.ie}
\end{center}

\begin{abstract}
\noindent The squared eigenfunctions of the spectral problem
associated with the Camassa–Holm (CH) equation represent a
complete basis of functions, which helps to describe the inverse
scattering transform for the CH hierarchy as a generalized Fourier
transform (GFT). All the fundamental properties of the CH
equation, such as the integrals of motion, the description of the
equations of the whole hierarchy, and their Hamiltonian
structures, can be naturally expressed using the completeness
relation and the recursion operator, whose eigenfunctions are the
squared solutions. Using the GFT, we explicitly describe some
members of the CH hierarchy, including integrable deformations for
the CH equation. We also show that solutions of some $(1+2)$ -
dimensional members of the CH hierarchy can be constructed using
results for the inverse scattering transform for the CH equation.
We give an example of the peakon solution of one such equation.

\end{abstract}

\vskip0.5cm


\vskip0.5cm

%
\section{Introduction}

\n The Camassa-Holm (CH) equation \cite{CH93} became famous as a
model in the theory of water waves. It is also known that it
describes axially symmetric waves in a hyperelastic rod
\cite{Dai98,CS2}. The most prominent representative of the
water-wave equations, the Korteweg-de Vries (KdV) equation does
not describe the wave-braking phenomenon. In addition to the
stable soliton solutions, the CH equation, together with another
recently derived nonlinear integrable equation, the
Degasperis-Procesi equation, has smooth solutions that develop
singularities in finite time via a process that captures the
features of the breaking waves: the solution remains bounded, but
the slope becomes unbounded \cite{CE98}. For the physical
relevance of these two equations as models for the propagation of
shallow water waves over a flat bottom one can consult, for
example \cite{J02,J03,DGH03,DGH04,I07,CL07,I-K07}. More about the
physical applications, modifications and the type of solutions of
the CH equation can also be found in \cite{CE98,
CS00,CS02,CM99,H05,I05,LO07,M06,M07,SS07}.

The CH equation has the form
\begin{equation}\label{eq1}
 u_{t}-u_{xxt}+2\omega u_{x}+3uu_{x}-2u_{x}u_{xx}-uu_{xxx}=0,
\end{equation}
where $\omega$ is a real constant. This equation is completely
integrable and admits a Lax pair \cite{CH93} \b \label{eq3}
\Psi_{xx}&=&\Big(\frac{1}{4}+\lambda (m+\omega)\Big)\Psi,
 \\\label{eq4}
\Psi_{t}&=&\Big(\frac{1}{2\lambda}-u\Big)\Psi_{x}+\frac{u_{x}}{2}\Psi+\gamma\Psi,
\e \n where $\gamma$ is an arbitrary constant and $m=u-u_{xx}$.
The CH solitary waves are stable solitons if $\omega
> 0$ \cite{J03, CS00, CS02, C03} or peakons if $\omega = 0$
\cite{CH93,BBS98,BBS99}.

The KdV and CH equations can also be interpreted as geodesic flow
equations for the respective $L^{2}$ and $H^{1}$ metrics on the
Bott-Virasoro group \cite{M98,CK03,CK06,K04,CI07,CKKT04}.

The CH equation is a bi-Hamiltonian equation, i.e. it admits two
compatible hamiltonian structures $J_1=(2\omega \partial
+m\partial+\partial m)$, $J_2=\partial-\partial^{3}$
\cite{CH93,FF81}:

\b m_t&=&-J_2\frac{\delta H_{2}[m]}{\delta m}=-J_1\frac{\delta
H_{1}[m]}{\delta m},\\ \label{H1a} H_1&=&\frac{1}{2}\int mu
\text{d}x,
\\\label{H2a} H_2&=&\frac{1}{2}\int (u^3+uu_x^2+2\omega u^2)
\text{d}x. \e

The infinite sequence of conservation laws (multi-Hamiltonian
structure) $H_n[m]$, $n=0,\pm1, \pm2,\ldots$, satisfying \b
J_2\frac{\delta H_{n}[m]}{\delta m}&=&J_1\frac{\delta
H_{n-1}[m]}{\delta m} \label{eq2a}\e can be computed explicitly
\cite{CH93,FS99,CI06,I06,CGI2}.

\section{Generalized Fourier transform}

The so-called recursion operator plays an important role in
describing an integrable hierarchy . The recursion operator for
the CH hierarchy is $ L=J_2^{-1}J_1$. The eigenfunctions of the
recursion operator are the squared eigenfunctions of the CH
spectral problem. For simplicity we consider the concrete case
where $m$ is a Schwartz class function, $\omega
>0$ and $m(x,0)+\omega > 0$. Then $m(x,t)+\omega
> 0$ for all $t$, e.g. see \cite{C01,H05}. It is convenient to introduce the notation: $q\equiv m+\omega$.
Let $k^{2}=-\frac{1}{4}-\lambda \omega$, i.e. \b \label{lambda}
\lambda(k)= -\frac{1}{\omega}\Big( k^{2}+\frac{1}{4}\Big).\e

\n A basis in the space of solutions of (\ref{eq3}) can be
introduced: $f^+(x,k)$ and $\bar{f}^+(x,\bar{k})$. For all real
$k\neq 0$ it is fixed by its asymptotic when $x\rightarrow\infty$
\cite{C01}, (also see \cite{CI06,CGI,ZMNP}): \b \label{eq6}
\lim_{x\to\infty }e^{-ikx} f^+(x,k)= 1, \e

\n   We can introduce another basis, $f^-(x,k)$
and$\bar{f}^-(x,\bar{k})$ fixed by its asymptotic when
$x\rightarrow -\infty$ for all real $k\neq 0$: \b \label{eq6'}
\lim_{x\to -\infty }e^{ikx} f^-(x,k)= 1, \e

\n Because $m(x)$ and $\omega$ are real we find that if $f^+(x,k)$
and $f^-(x,k)$ are solutions of (\ref{eq3}) then
\begin{equation}\label{eq:inv}
 \bar{f}^+(x,\bar{k}) = f^+(x,-k), \qquad \mbox{and} \qquad
 \bar{f}^-(x,\bar{k}) = f^-(x,-k),
\end{equation}
are also solutions of (\ref{eq3}). The squared solutions are
\begin{equation}\label{eq23} F^{\pm}(x,k)\equiv (f^{\pm}(x,k))^2, \qquad F_n^{\pm}(x)\equiv
F(x,i\kappa_n),
\end{equation}

\n where $F_n^{\pm}(x)$ are related to the discrete spectrum
$k=i\kappa_n$, where $ 0<\kappa_1<\ldots<\kappa_n<1/2$.

\n Using the asymptotics (\ref{eq6}), (\ref{eq6'}) and the Lax
equation (\ref{eq3}) one can show that
\begin{equation}\label{eq23a} L_{\pm}F^{\pm}(x,k)=\frac{1}{\lambda}F^{\pm}(x,k).
\end{equation}

\n where \b L_{\pm}=(\partial^2-1)^{-1}\Big[4q(x)-2\int_{\pm
\infty}^{x}\text{d}\tilde{x} \, m'(\tilde{x})\Big]\label{eq44}\e
is the recursion operator. The inverse of this operator is also
well defined.

We introduce the notation $\partial_{\pm}^{-1}\equiv \int_{\pm
\infty}^x{\text d}\tilde{x}$. The squared solutions (\ref{eq23})
form a complete basis in the space of the Schwartz class functions
$m(x)$, and $y$, $t$, can be treated as some additional
parameters. Also, the Generalised Fourier Transform (GFT) for $q$
and its variation over this basis is \cite{CGI2}
\begin{equation}\label{Compl1} \sqrt{\frac{\omega}{q(x)}}-1=
\pm\frac{1}{2\pi i}\int_{-\infty}^{\infty}\frac{2k
\mathcal{R}^{\pm}(k)} {\omega \lambda(k)} F^{\pm}(x,k)\text{d}k +
\sum_{n=1}^{N}\frac{2\kappa_n}{\omega
\lambda_n}R_n^{\pm}F_n^{\pm}(x),
\end{equation}
\b \frac{\partial_{\pm}^{-1}\delta(\sqrt{q})}{\sqrt{q}}&=&
\frac{1}{2\pi i}\int_{-\infty}^{\infty}\frac{i\delta
\mathcal{R}^{\pm}(k)}{\omega
\lambda(k)}F^{\pm}(x,k)\text{d}k \nonumber  \\
& \pm & \sum_{n=1}^{N}\Big[\frac{\delta R_n^{\pm}-R_n^{\pm}\delta
\lambda_n }{\omega \lambda_n}F_n^{\pm}(x)
+\frac{R_n^{\pm}}{i\omega\lambda_n}\delta \kappa_n
\tilde{F}_n^{\pm}(x) \Big], \label{Compl2}  \e

\n where $\tilde{F}_n^{\pm}(x)\equiv \frac{\partial}{\partial
k}F^{\pm}(x,k)|_{k=i\kappa_n}$. The generalized Fourier
coefficients $\mathcal{R}^{\pm}(k)$, $R_n^{\pm}$, together with
the set of discrete eigenvalues, are called {\it scattering data}.
The variation is with respect to any additional parameter, e.g.
$y$, $t$.

The equations of the CH Hierarchy can be written as \b
P_2(L_{\pm})\frac{2\partial_{\pm}^{-1}(\sqrt{q})_t}{\sqrt{q}}+P_1(L_{\pm})\big(\sqrt{\frac{\omega}{q}}-1\Big)=0,\label{CHH}\e

\n where $P_1(z)$ and $P_2(z)$ are two polynomials. If
$\Omega(z)=\frac{P_1(z)}{P_2(z)}$ is a ratio of these two
polynomials one can define $\Omega(L_{\pm})\equiv
P_1(L_{\pm})P_2^{-1}(L_{\pm})$ (provided $P_2(L_{\pm})$ is an
invertible operator). Then (\ref{CHH}) can be written in the
equivalent form \b q_t+2q\tilde{u}_x+q_x\tilde{u}=0,\qquad
\tilde{u}=\frac{1}{2}\Omega(L_{\pm})\Big(\sqrt{\frac{\omega}{q}}-1\Big).\label{eqCHH}
\e

Due to the completeness of the squared eigenfunctions basis, from
(\ref{CHH}), (\ref{Compl1}) and (\ref{Compl2}) we obtain linear
differential equations for the scattering data: \b
\mathcal{R}^{\pm}_t \mp
ik\Omega(\lambda^{-1})\mathcal{R}^{\pm}(k)=0, \\ R^{\pm}_{n,t}
\pm \kappa_n \Omega(\lambda_n^{-1}) R^{\pm}_n=0, \\
\lambda_{n,t}=0. \e

The GFT for other integrable systems is derived e.g. in
\cite{K76,GeHr1,G86,GI92,GY94,IKK94}.

{\bf Example:} We now consider the case
$\Omega(z)=a_{-1}z^{-1}+a_0+a_1z$ (where $a_j$ are constants). The
(\ref{CHH}) equation can then be rewritten as \b
L_{\pm}\frac{2\partial_{\pm}^{-1}(\sqrt{q})_t}{\sqrt{q}}+(a_{-1}+a_0L_{\pm}+a_1L_{\pm}^2)\Big(\sqrt{\frac{\omega}{q}}-1\Big)=0,\label{E1}\e

\n Taking the identities \b
L_{\pm}\frac{2\partial_{\pm}^{-1}(\sqrt{q})_t}{\sqrt{q}}&=&-4\partial_{\pm}^{-1}
u_t, \qquad
\frac{1}{2}L_{\pm}\Big(\sqrt{\frac{\omega}{q}}-1\Big)=u, \nonumber
\\ L_{\pm}u&=&-2(1-\partial^2)^{-1}\partial_{\pm}^{-1}(uq_x+2qu_x)
\nonumber \e into account, we obtain an integrable equation \b
q_t+a_1(2qu_x+q_xu)-\frac{a_0}{2}q_x-\frac{a_{-1}}{4}(\partial-\partial^3)\sqrt{\frac{\omega}{q}}=0,\label{E2}
\e

\n which becomes the Camassa-Holm equation (\ref{eq1}) with the
choice $a_1=1$, $a_0=a_{-1}=0$. Therefore, (\ref{E2}) can be
considered as an integrable 'deformed' version of the CH equation.
Another choice for the constants, $a_{-1}=1$, $a_0=a_1=0$, leads
to the extended Dym equation \cite{CH93,CGI2,FOR96}. If $a_1=1$,
$a_{-1}=0$ but $a_0\ne 0$ the equation is usually called
Dullin-Gottwald-Holm Equation \cite{DGH03,DGH04,M06,M07}.

The Hamiltonian of (\ref{eqCHH}) with respect to the Poisson
bracket related to the Hamiltonian operator $J_1$,
\begin{equation}\label{PBJ1}
\{A,B\}=-\int (\omega+m)\Big(\frac{\delta A}{\delta m}\partial
\frac{\delta B}{\delta m}- \frac{\delta B}{\delta m}\partial
\frac{\delta A}{\delta
 m}\Big)\text{d}x,
\end{equation}

\n is (see \cite{CGI}) \b H^{\Omega}=\int_{0}^{\infty} \frac{k^2
\Omega(\lambda^{-1})}{\pi \omega \lambda(k)^2}
\ln\Big(1-\mathcal{R}^{\pm}(k)\mathcal{R}^{\pm}(-k)\Big)\text{d}k
-\frac{2}{\omega}\sum_{n=1}^{N}\int \frac{\kappa_n
^2}{\lambda_n^2} \Omega(\lambda_n^{-1}) \text{d}
\kappa_n.\label{eq84x} \e

In (\ref{eq84x}) the Hamiltonian is given in terms of the
Scattering data. In general, it is not straightforward to find the
corresponding expressions in terms of the field variable $q(x)$
(or $m(x)$). For example, the Hamiltonian of (\ref{E2}) with
respect to (\ref{PBJ1}) is \b
H^{\Omega}=a_1H^{CH}_1-\frac{a_0}{2}I_0-
\frac{a_{-1}}{4}H^{CH}_{-1},\nonumber \e

\n where $H_1^{CH}=\frac{1}{2}\int_{-\infty}^{\infty} mu {\text d}
x$ is the first CH Hamiltonian (\ref{H1a}),

\begin{equation}
H^{CH}_{-1}=\frac{1}{2}\int_{-\infty}^{\infty}\Big[\Big(\sqrt[4]{\frac{\omega}{q}}-\sqrt[4]{\frac{q}{\omega}}\Big)^2+\frac{\sqrt{\omega}q_x^2}{4q^{5/2}}
\Big]\text{d}x,\label{eq90}\end{equation}

\n is the (-1)-st Hamiltonian for the CH equation, and the
integral \b I_0&=&\int_{-\infty}^{\infty} m {\text
d}x=H_0^{CH}+2\omega \alpha, \label{I_0} \e

\n is related to the other two CH integrals \cite{CGI2} \b
H_0^{CH}=\int_{-\infty}^{\infty} (\sqrt{q}- \sqrt{\omega})^2
{\text d} x, \qquad \alpha=\int_{-\infty}^{\infty}
\Big(\sqrt{\frac{\omega}{q}}-1\Big){\text d} x.\nonumber \e

\section{Peakon solutions of a (1+2) - dimensional equation from the
CH hierarchy}

We consider an integrable member of the CH hierarchy with two
'time' variables - $t$ and $y$ (cf. \cite{CGP}) \b
q_t+2(U_{xy}+\eta U_{xx})q+(U_y+\eta U_x+\gamma)q_x=0, \qquad
q=U_x-U_{xxx}+\omega, \label{CHy} \e where $\omega$, $\gamma$ and
$\eta$ are arbitrary constants. The Lax pair for (\ref{CHy}) is

\b \Psi_{xx}&=&\Big(\frac{1}{4}+\lambda (m+\omega)\Big)\Psi,
 \nonumber \\
\Big(\partial_t-\frac{1}{2\lambda}\partial_y\Big)\Psi&=&-\Big(U_y+\eta
U_x+\gamma -\frac{\eta}{2\lambda}
\Big)\Psi_{x}+\frac{1}{2}(U_{xy}+\eta U_{xx})\Psi. \nonumber \e

\n The Lax pair represents a non-isospectral problem. Indeed, the
equation (\ref{CHy}) can be written as a compatibility condition
\b
\Big(\partial_t-\frac{1}{2\lambda}\partial_y\Big)(\Psi_{xx})=\partial_x^2\Big(\partial_t-\frac{1}{2\lambda}\partial_y\Big)\Psi,
\nonumber \e where $\lambda$ satisfies the relaxed condition
$\lambda_t-\frac{1}{2\lambda}\lambda_y=0$. But we assume that the
spectrum is $t$- and $y$-independent in what follows, and we can
generalise the solutions obtained for the CH equation.

\n We can also write (\ref{CHy}) as \b (\sqrt{q})_t+[(U_y+\eta
U_x+\gamma)\sqrt{q}]_x=0.\e  Then \b
\partial_{\pm}^{-1}(\sqrt{q})_t+(U_y+\eta
U_x+\gamma)\sqrt{q}+\beta=0,\e where $\beta$ is an integration
constant. Further, choosing $\beta=-\gamma\sqrt{\omega}$ and using
the identities \b U_y=-\frac{1}{2}L_{\pm}
\Big(\frac{\partial_{\pm}^{-1}(\sqrt{q})_y}{\sqrt{q}}\Big),\qquad
U_x=\frac{1}{2}L_{\pm} \Big(\sqrt{\frac{\omega}{q}}-1\Big) \e

\n we can write (\ref{CHy}) in the form \b
\frac{\partial_{\pm}^{-1}(\sqrt{q})_t}{\sqrt{q}}-
\frac{1}{2}L_{\pm}
\Big(\frac{\partial_{\pm}^{-1}(\sqrt{q})_y}{\sqrt{q}}\Big)+\Big(\frac{\eta}{2}L_{\pm}-\gamma\Big)
\Big(\sqrt{\frac{\omega}{q}}-1\Big)=0. \label{2d}\e

\n Taking  (\ref{2d}), (\ref{Compl1}) and (\ref{Compl2}) into
account and considering variations with respect to $y$ and $t$ we
obtain linear equations for the scattering data: \b
\mathcal{R}^{\pm}_{t} -\frac{1}{2\lambda}\mathcal{R}^{\pm}_{y}\pm
2ik\Big(\gamma-\frac{\eta}{2\lambda}\Big) \mathcal{R}^{\pm}=0,
\\R^{\pm}_{n,t}
-\frac{1}{2\lambda_n}R^{\pm}_{n,y}\mp2\Big(\gamma-\frac{\eta}{2\lambda_n}\Big)\kappa_n
R^{\pm}_n=0, \e

\n assuming $ \lambda_{n,t}=0$. For example, if $\gamma=\eta=0$,
then the solution is any function of $t+2\lambda y$ (with
appropriate decaying properties): \b
\mathcal{R}^{\pm}(y,t)=\mathcal{R}^{\pm}(t+2\lambda y), \qquad
R^{\pm}_{n}(y,t)=R^{\pm}_{n}(t+2\lambda_ny). \e

\n We can obtain CH equation itself (\ref{eq1}) for $x=y$,
$u=U_x$, $\gamma=\eta=0$.

We demonstrate how to write explicit peakon solutions for this
equation ($\gamma=\eta=\omega=0$). Until now, $\omega$ was
strictly positive in our considerations, but we can take the limit
$\omega\rightarrow 0$ \cite{dMS} which produces 'peaked' solitons
and the equations for the scattering data should also hold in this
case. We assume that $x_k=x_k(t,y)$, $p_k=p_k(t,y)$ and introduce
the notations $\varepsilon(x)\equiv \text{sign}(x)$,
$p'_k(t,y)=\frac{\partial p_k}{\partial y}$ etc. The ansatz that
produces the $N$-peakon solution for the CH equation can be
generalised as \b q(x;t,y)=\sum_{k=1}^N p_k \delta(x-x_k).
\nonumber \e We hence obtain \b U(x;t,y)&=&\frac{1}{2}\sum_{k=1}^N
p_k\varepsilon(x-x_k)(1-e^{-|x-x_k|}), \nonumber \\
U_y(x;t,y)&=&\frac{1}{2}\sum_{k=1}^N[
p'_k\varepsilon(x-x_k)(1-e^{-|x-x_k|})-p_k e^{-|x-x_k|}x'_k],
\nonumber \\
U_{xy}(x;t,y)&=&\frac{1}{2}\sum_{k=1}^N[ p'_k e^{-|x-x_k|}+p_k
\varepsilon(x-x_k) e^{-|x-x_k|}x'_k]. \nonumber \e

\n Using this ansatz and the identity \b
f(x)\delta'(x-x_0)=f(x_0)\delta'(x-x_0)-f'(x_0)\delta(x-x_0)\nonumber
\e we obtain the following system of PDEs for the quantities
$x_k(t,y)$, $p_k(t,y)$ from (\ref{CHy}): \b
\dot{x}_l&=&\frac{1}{2}\sum_{k=1}^N[
p'_k\varepsilon(x_l-x_k)(1-e^{-|x_l-x_k|})-p_k
e^{-|x_l-x_k|}x'_k],
\label{peak1} \\
\dot{p}_l&=&-\frac{1}{2}p_l\sum_{k=1}^N[ p'_k e^{-|x_l-x_k|}+p_k
\varepsilon(x_l-x_k) e^{-|x_l-x_k|}x'_k], \label{peak2} \e where
$\dot{x}_k(t,y)=\frac{\partial x_k}{\partial t}$ etc. The
solutions of this system can be obtained from the $N$-peakon
solution for the CH equation \cite{BBS98, BBS99}, where the
scattering data are now arbitrary functions of their argument
(with appropriate decaying properties): $R_n \equiv
R^{+}_{n}(t+2\lambda_ny)$. For example, if $N=1$, then the system
has the form \b \dot{x}_1+\frac{1}{2}p_1 x'_1=0, \qquad
\dot{p}_1+\frac{1}{2}p_1 p'_1=0, \nonumber \e with a solution
$p_1=-\frac{1}{\lambda_1}=\text{const}$, $x_1=\ln R_1(t+2\lambda_1
y)$, cf. \cite{BBS98, BBS99}. The system for $N=2$ peakons is (we
assume that $x_1<x_2$ for all $y$ and $t$, i.e. the case for which
the $N$-peakon solution for the CH equation is obtained in
\cite{BBS98, BBS99}) \b
\dot{x}_1&=&\frac{1}{2}[-p'_2(1-e^{-|x_1-x_2|})-p_1x'_1-p_2
e^{-|x_1-x_2|}x'_2],
\label{x1} \\
\dot{x}_2&=&\frac{1}{2}[p'_1(1-e^{-|x_1-x_2|})-p_1e^{-|x_1-x_2|}x'_1-p_2
x'_2], \label{x2} \\
\dot{p}_1&=&-\frac{1}{2}p_1[ p'_1+p'_2 e^{-|x_1-x_2|}-p_2
e^{-|x_1-x_2|}x'_2], \label{p1} \\
\dot{p}_2&=&-\frac{1}{2}p_2[ p'_1 e^{-|x_1-x_2|}+p'_2+p_1
e^{-|x_1-x_2|}x'_1], \label{p2} \e

\n with solutions \cite{BBS98, BBS99} \b p_1&=&-\frac{\lambda_1^2
R_1+\lambda_2^2 R_2}{\lambda_1\lambda_2(\lambda_1 R_1+\lambda_2
R_2)}, \qquad
p_2=-\frac{R_1+R_2}{\lambda_1 R_1+\lambda_2 R_2} \nonumber \\
x_1&= & \ln \frac{(\lambda_1-\lambda_2)^2R_1R_2}{\lambda_1^2
R_1+\lambda_2^2 R_2}, \qquad x_2=\ln (R_1+R_2), \nonumber  \e

\n where $R_k=R_k(t+2\lambda_k y)$ (this can be easily verified).
We note that the total momentum
$p_1+p_2=-\frac{\lambda_1+\lambda_2}{\lambda_1\lambda_2}$ is
conserved. Of course, in general, for an arbitrary $N$, the
$N$-peakon solution for the CH equation obtained in \cite{BBS98,
BBS99} can be used because the inverse scattering method for
hierarchy (\ref{eqCHH}) is the same as for the CH equation
\cite{CGI2,CGI}. The only difference is the time dependence of the
scattering data (and/or the additional $y$-dependence, etc.). In
this example, $y$ has the meaning of a second 'time' variable.
Clearly, the conserved quantities in terms of $x_k$ and $p_k$ have
the same form as those for the CH peakons and can be expressed in
terms of the quantities $\lambda_k$, which we already assumed to
be independent of $y$ and $t$. But the Hamiltonian formulation is
problematic because formally $\Omega(z)\equiv 0$ for the peakon
solution and the Hamiltonian with respect to (\ref{PBJ1}) is
degenerate because of (\ref{eq84x}). Moreover, the right-hand side
of system (\ref{x1}) -- (\ref{p2}) involves not only the
quantities $x_k$ and $p_k$ but also their $y$-derivatives.

The explicit dependence on the scattering data given in \cite{CGI}
can be used in the same way for the $N$-soliton solutions of the
CH hierarchy. The situation where the initial data condition
$q(x,0)\equiv m(x,0)+\omega>0$ does not hold is more complicated
and requires a separate analysis \cite{CE98,C01,K05}.

$\phantom{*}$

{ \bf Acknowledgments}. The author is thankful to Prof. A.
Constantin for the opportunity to visit Lund University (Sweden)
where part of the research presented in this work has been done.

This work was supported by the G. Gustafsson Foundation for
Research in Natural Sciences and Medicine (Sweden).

\label{lastpage}
\end{document}